\documentclass[twocolumn,superscriptaddress,floatfix,prl,showpacs]{revtex4-1}
\usepackage{graphicx}
\usepackage{amsmath}
\usepackage{amssymb}
\usepackage{color}
\usepackage{braket}
\usepackage{dsfont}

\begin{document}
\title{Topological quantum transition driven by charge-phonon coupling in the Haldane Chern insulator}

\author{L.~M.~Cangemi}
\affiliation{SPIN-CNR and Dip. di Scienze Fisiche - Universit\`a di Napoli Federico II - I-80126 Napoli, Italy}

\author{A.~S.~Mishchenko}
\affiliation{RIKEN Center for Emergent Matter Science (CEMS),  
2-1 Hirosawa, Wako, Saitama, 351-0198, Japan}
\affiliation{NRC ``Kurchatov Institute", 123182, Moscow, Russia}

\author{N.~Nagaosa}
\affiliation{RIKEN Center for Emergent Matter Science (CEMS),
2-1 Hirosawa, Wako, Saitama, 351-0198, Japan}
\affiliation{Department of Applied Physics and Quantum-Phase Electronics Center, University of Tokyo, Bunkyo, Tokyo 113-8656, Japan} 

\author{V. Cataudella}
\affiliation{SPIN-CNR and Dip. di Scienze Fisiche - Universit\`a di Napoli Federico II - I-80126 Napoli, Italy}

\author{G. De Filippis}
\affiliation{SPIN-CNR and Dip. di Scienze Fisiche - Universit\`a di Napoli Federico II - I-80126 Napoli, Italy}

\begin{abstract}
In condensed matter physics many features can be understood in terms of their topological properties. Here we report evidence of a topological quantum transition driven by the charge-phonon coupling in the spinless Haldane model on a honeycomb lattice, a well-known prototypical model of Chern insulator. Starting from parameters describing the topological phase in the bare Haldane model, we show that the increasing of the strength of the charge lattice coupling drives the system towards a trivial insulator. The average number of fermions in the Dirac point, characterized by the lowest gap, exhibits a finite discontinuity at the transition point and  can be used as direct indicator of the topological quantum transition. Numerical simulations show, also, that the renormalized phonon propagator exhibits a two peak structure across the quantum transition, whereas, in absence of the mass term in the bare Hadane model, there is indication of a complete softening of the effective vibrational mode signaling a charge density wave instability.
\end{abstract}
\maketitle

In the last decades, topological insulators have seen a tremendous growth of attention: starting from the early-days discovery of the integer Quantum Hall Effect (QHE) \cite{Von,Th}, and fractional QHE \cite{Tsui,Lau}, which is now considered a prototype of topologically ordered state \cite{Hasan}, the field has achieved a widespread popularity following the seminal works by Kane, Mele \cite{Kane}, and Bernevig et al. \cite{ber}, which predicted the occurrence of time-reversal-symmetry protected topological phases in 2D, known as Quantum Spin Hall Effect \cite{hal1,moore}. Great amount of work has been done to extend these theories to 3D \cite{Hasan,Qi}. Several distinct groups of candidate materials have been proposed for the observation of these novel phases.

A wide class of theories describing topological insulators, i.e. simmetry-protected topological insulators, are based on free-fermion models, with topologically non-trivial band structures \cite{ber1}. These phases have been classified according to a set of rules \cite{kit,sch,jef}. Chern insulators belong to the family of QHE, i.e. symmetry class $A$, are based on free fermion theories, and allow the occurrence of chiral edge-modes \cite{jac}. Due to these properties, Chern insulating phase has been considered as the simplest meaningful example of topological phase \cite{rac}. The spinless Haldane model on a honeycomb lattice is a well-known prototypical model of Chern insulator \cite{hal}. It is a paradigmatic example of a Hamiltonian featuring topologically distinct phases, where the quantum Hall effect appears as an intrinsic property of a band structure, rather than being caused by an external magnetic field. This model has recently regained attention, following its experimental realization using cold atoms platforms \cite{jot}, as well as interacting superconducting circuits \cite{rou}.

In general, so far the most of theoretical work has been done aimed at understanding the effect of  Coulomb correlations on the topological properties \cite{1,2,3,4,5,naoto,prok}. 
On the other hand, electron-phonon interaction is so inevitably present in any solid that, from the first principles, one cannot even distinguish and separate Coulomb and electron-phonon interaction because they are unambiguously connected \cite{Tupi}.
The issue of Coulomb correlations is so well developed that the  focus of current studies is already the settling of the details of the already known phase diagrams using better and better methods, see e.g. \cite{prok}. To the contrary, there are only few studies of the influence of electron-phonon coupling (EPC) \cite{mona,hol} all considering  models different from Haldane Chern insulator. To fill this gap, we account for the lattice quantum dynamics including on-site optical phonons coupled {\it a la} Holstein to spinless fermions, described by the Haldane model. We perform a numerical study of the bulk properties of the interacting system, employing Cluster Pertubation Theory(CPT) \cite{sen}, that, starting from the exact numerical computation of the the Green function, performed on a suitably chosen cluster, allows to compute the interacting Green functions of the whole lattice, an experimentally accessible function through angle resolved photoemission spectroscopy measurements \cite{damascelli}.

We find evidence of a topological phase transition driven by EPC. Starting from the topological phase in the bare Haldane model, we show that the increasing of the strength of the EPC drives the system towards a trivial insulator. Across the phase transition, a strong hybridization of the quasiparticle bands of the bare Haldane model occurs. Numerical simulations show also that the renormalized phonon propagator exhibits a two peak structure across the quantum transition, whereas, in absence of the mass term, there is indication of a complete softening of the effective vibrational mode signaling a charge density wave instability.

{\it The model.}
The Haldane model \cite{hal} describes spinless fermions on a honeycomb lattice at half-filling. It includes an on-site mass term, $M$, and a complex next nearest neighbor hopping. $M$ breaks the inversion simmetry of the lattice. The complex tunnelling breaks the time-reversal simmetry, realizing a staggered magnetic field on the lattice without a net magnetic flux through the plaquette. The Hamiltonian reads
\begin{equation}\label{eq:Haldane}
  H_{H}=-\sum_{i,j}t_{i,j} c_{i}^{\dagger}c_{j}+ M\sum_{i}\xi_{i}c_{i}^{\dagger}c_{i}
\end{equation}
where $c_{i}^{\dagger}(c_{i})$ are fermionic creation (annihilation) operators on the site $i$, $t_{i,j}=t_{1}$ ($t_2 e^{i\xi_{i}\phi}$) is the nearest (next nearest) neighbor electronic hopping, and $\xi_{i}$ is an integer which takes the values $\pm 1$ respectively on the two sublattices $(A,B)$. The model in Eq.(\ref{eq:Haldane}) describes a topological Chern insulator, gapped at Dirac points $\text{K}\mbox{, } \text{K}^{\prime}$ of the Brillouin zone. By rewriting Eq.(\ref{eq:Haldane}) in the quasi-momentum space, and taking as $(a_i,b_i)$ the two different fermionic operators at each sublattice site, it can be found:
\begin{equation}
  H_{H}=\sum_k
  \begin{pmatrix}a^{\dagger}_k & b^{\dagger}_k
  \end{pmatrix}
  \mathcal{H}(\boldsymbol{k})
  \begin{pmatrix} a_k\\b_k
  \end{pmatrix}
\end{equation}
with $\mathcal{H}(\boldsymbol{k})=\epsilon(\boldsymbol{k})\, {\mathds{1}} + \boldsymbol{h(\boldsymbol{k})}\cdot{\boldsymbol{\sigma}}$. Here $\boldsymbol{\sigma}=\sigma_i\mbox{, } i=x,y,z$ denote Pauli matrices, the quasi-momentum $\boldsymbol{k}=(k_x\mbox{, }k_y)$ belongs to the first Brillouin zone, and $\boldsymbol{h}(\boldsymbol{k})$ is a 3D vector, $\boldsymbol{h}(\boldsymbol{k})=(h_x(\boldsymbol{k}),h_y(\boldsymbol{k}),h_z(\boldsymbol{k}))$ \cite{Supplement}. Units are such that $\hbar=1$. In order to diagonalize this two band model, we have to perform an unitary transformation and, then, introduce a new set of wavefunctions: $|{k,+} \rangle$, and $|{k,-}\rangle$ \cite{Supplement}. The corresponding creation and annihilation operators, $\gamma^{\dagger}_{k,\pm}$ and $\gamma_{k,\pm}$, define the quasiparticles of the Haldane model: $H_{H}=\sum_k (E_{k,-} \gamma^{\dagger}_{k,-} \gamma_{k,-}+ E_{k,+} \gamma^{\dagger}_{k,+} \gamma_{k,+})$. Here $E_{k,\mp}$ represent the lower and upper energy bands with respect to the chemical potential $\mu$. The gap in the points $\text{K}\mbox{, } \text{K}^{\prime}$ is given, respectively, by:
\begin{equation}\label{eq:gap}
  \Delta =2 \mid M\mp 3\sqrt{3}t_2\sin\phi\mid.
\end{equation}
The Haldane model predicts the existence of different parameter regions in which the system behaves as an insulator, separated by a curve in the parameter space where gap closure occurs. These two regions describe topologically distinct insulating phases, marked by the values of a topological invariant, i.e. the Chern number $C_h$. The topologically non-trivial insulating phase is characterized by $C_h=\pm 1$, and it occurs for $ -M_c< M < M_c$, where $M_c \equiv 3\sqrt{3}t_2\sin\phi$. In all other cases $C_h=0$, and the system behaves as a trivial insulator.

Our aim is to investigate the effects on the topological properties of the system when fermions couple to the lattice degrees of freedom. We introduce in the Hamiltonian Eq.(\ref{eq:Haldane}) an interaction term {\it a la} Holstein, where charge fluctuations are linearly coupled to the displacement of local lattice vibrations:
\begin{equation}\label{eq:Interacting}
  H=H_{H} + \omega_{0}\sum_i d^{\dagger}_i d_i + g\omega_{0}\sum_i(c^{\dagger}_i c_i-\frac{1}{2})(d^{\dagger}_i + d_i)
\end{equation}
We employ shorthand notation $d^{\dagger}_i$($d_i$) for two different bosonic operators, which respectively create (annihilate) a phonon on the two $\left ( A,B \right)$ sublattice sites, $\omega_{0}$ is the optical mode frequency, and  $g$ represents the strength of the coupling with lattice. We introduce also the dimensionless parameter $\lambda=g^2 \omega_0/4 t_1$. Here we restrict our attention to the case of half-filling, i.e. $\sum_{i \in A} a^{\dagger}_i a_i + \sum_{i \in B} b^{\dagger}_i b_i =N_c=N/2$, where $N_c$ ($N$) is the number of unit cells (lattice sites).

{\it Lang-Firsov approach (LFA).}
If the optical mode frequency is the highest energy scale (antiadiabatic regime), i.e. $\omega_0 \gg t_1, t_2, M$, the physics is well captured by LFA \cite{mahan}, that is based on the unitary transformation: $\tilde{H}=e^{S} H e^{-S}$, where $ S = g \sum_i (c^{\dagger}_i c_i -\frac{1}{2}) (d^{\dagger}_i - d_i)$. In the new basis, the electronic hopping is assisted by phononic operators that, in the antiadiabatic regime, can be treated as a small perturbation. This approximation leads to renormalized values of $t_1$ and $t_2$ through the factor $e^{-\frac{4 \lambda t_1}{\omega_0}}$. On the other hand, as it is straightforward to verify, the value of $M$ is not affected by the unitary transformation. The net result is that it is possible to replace the Hamiltonian $\tilde{H}$ with that of an effective Haldane model, where now the parameters, and then the topological-trivial insulator transition, are controlled by the strength of the EPC. In other words, by increasing the value of $\lambda$, it is possible to induce a topological quantum transition. On the other hand, this approach becomes exact only in the limit $t_1=t_2=0$. In order to investigate if these effects survive for parameter values of physical interest, a more accurate treatment of EPC is needed. To this aim we employ the CPT \cite{Supplement}, that allows us to compute the electronic Green function, $G_{i,j}(\boldsymbol{q},z)$, of the interacting system, from which detailed informations on the renormalized band structure as well as spectral functions can be derived. Here $i$ stands for $\left ( A,B \right )$, i.e. indicates the two sublattices, and $z=\omega+i\eta$ lies in the complex upper half plane. Starting from $G_{i,j}$, it is straightforward to derive the Green functions $G_{(+,+)}$ and $G_{(-,-)}$, corresponding to the quasiparticle operators of the bare Haldane model. We will focus our attention on the following set of parameters: $t_2/t_1=0.3$, $\omega_0=3 t_1$, $\phi=\frac{\pi}{2}$ and two different values of $M$, i.e. $M_1=0.94 M_c$ and $M_2=0.42 M_c$. In the absence of EPC, these two values describe the topological insulator phase near and far from, respectively, the transition towards to a trivial insulator. Furthermore, for these values of the parameters, the lowest gap is located at $\text{K}$ point, and $H_H$ exhibits hole-particle symmetry so that $\mu=0$.     

\begin{figure}[thb]
\includegraphics[width=0.99\columnwidth]{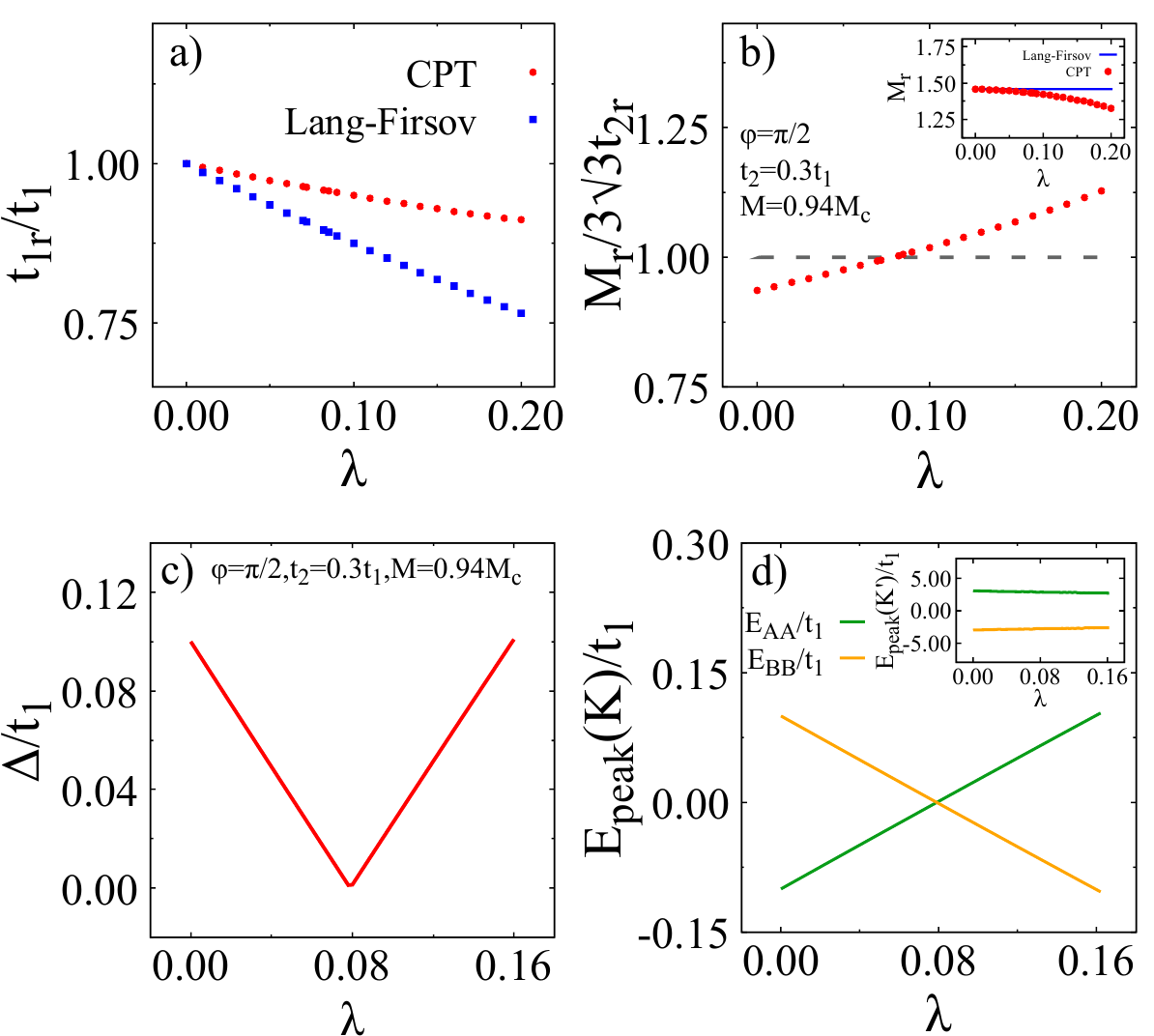}
\caption{\label{fig:1} (color online) 
  (a) and (b): parameters of the effective Haldane model vs $\lambda$; (c) and (d): behavior of the gap and the energies of the peaks of $A_{(A,A)}$ and $A_{(B,B)}$, at $\text{K}$ and $\text{K}^{\prime}$, as function of $\lambda$.  
}
\end{figure}

{\it The results.}
Within the hole sector, near $\text{K}$ point, we followed the dispersion of lowest energy quasiparticle peak associated to one of the two spectral weight functions $A_{(\mp,\mp)}(\boldsymbol{q},\omega)=-\frac{\Im{G_{(\mp,\mp)}(\boldsymbol{q},z)}}{\pi}$. It turns out to be equivalent to that of an effective Haldane model. In Fig.~\ref{fig:1}a and Fig.~\ref{fig:1}b we plot, as function of $\lambda$, the renormalized values of the electronic hopping and $M$, i.e. $t_{1r}$, $t_{2r}$ and $M_r$, and compare them with those predicted within LFA. In the CPT all the parameters, including $M$, are renormalized, but, also in this approach, a topological quantum transition occurs. Indeed, around $\lambda_c \simeq 0.08$, the ratio $\frac{M_r}{3\sqrt{3}t_{2r}}$ becomes greater than 1, signaling the phase transition. Fig.~\ref{fig:1}c shows that the gap, by increasing $\lambda$, first decreases, at $\lambda_c$ becomes zero, and then increases. It is also worth noting that, within the bare Haldane model, the spectral weight functions corresponding to the two sublattices assume the following form: $A_{(A,A)}(\boldsymbol{q},\omega)=\frac{(1+n_z)}{2}\delta(\omega-E_{q,+})+\frac{(1-n_z)}{2}\delta(\omega-E_{q,-})$ and $A_{(B,B)}(\boldsymbol{q},\omega)=\frac{(1-n_z)}{2}\delta(\omega-E_{q,+})+\frac{(1+n_z)}{2}\delta(\omega-E_{q,-})$,  where $n_z=\frac{h_z}{\parallel\boldsymbol{h}\parallel}$. Here $\delta(\omega)$ is the Dirac delta function. On the other hand, we emphasize that, again at $\lambda=0$, if the system is in the topological phase $0<M<M_c$, $n_z$, when evaluated at $\text{K}$ and $\text{K}^{\prime}$, assumes opposite values, respectively $-1$ and $1$, whereas, in the trivial insulating phase ($M>M_c$), $n_z=1$ at both $\text{K}$ and $\text{K}^{\prime}$. Then, at $\lambda=0$, it is clear that $A_{(A,A)}(\text{K},\omega)$ ($A_{(B,B)}(\text{K},\omega)$) is peaked only at $E_{\text{K},-}$ ($E_{\text{K},+}$) in the topological phase and only at $E_{\text{K},+}$ ($E_{\text{K},-}$) in the trivial insulating phase. On the other hand, $A_{(A,A)}(\text{K}^{\prime},\omega)$ ($A_{(B,B)}(\text{K}^{\prime},\omega)$) has spectral weight different from zero only at $E_{\text{K}^{\prime},+}$ ($E_{\text{K}^{\prime},-}$), independently on the phase. We followed, as function of $\lambda$, the peak position of these two spectral functions at both $\text{K}$ and $\text{K}^{\prime}$. Fig.~\ref{fig:1}d shows that at $\lambda<\lambda_c$ ($\lambda>\lambda_c$) the behavior of the fermions on the two sublattices is in agreement with that predicted by the bare Haldane model in the topological (trivial) insulating phase. It confirms that at $\lambda_c$ a quantum transition occurs.

\begin{figure}[thb]
\includegraphics[width=0.99\columnwidth]{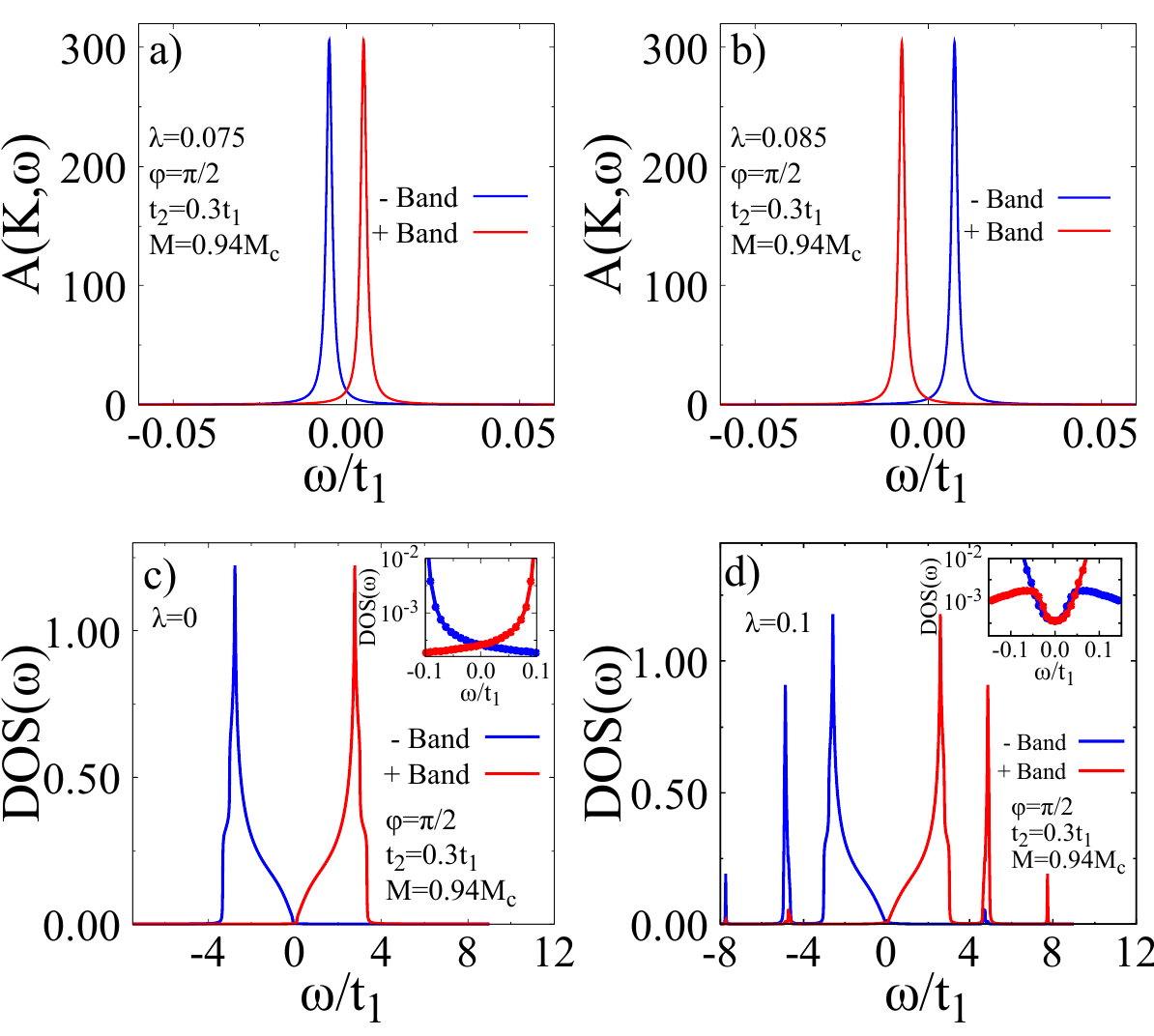}
\caption{\label{fig:2} (color online)
  (a) and (b): $A_{(-,-)}$ and $A_{(+,+)}$, at $\text{K}$, just below and above $\lambda_c$;
  (c) and (d): density of states with zoom (insets) around $\mu$ ($\omega=0$), at $\lambda=0$ and $\lambda>\lambda_c$.
}  
\end{figure}

Now we focus our attention on the spectral weight functions corresponding to the operators describing the quasiparticles in the bare Haldane model, i.e. $A_{(-,-)}$ and $A_{(+,+)}$. These two functions, at the Dirac point $\text{K}$, are plotted in Fig.~\ref{fig:2}a and Fig.~\ref{fig:2}b for two different values of $\lambda$, $\lambda=0.075$ and $\lambda=0.085$, respectively before and after the topological phase transition. Crossing $\lambda_c$, the energy gap closes and opens again, and, at the same time, the character of the two bands changes, i.e. the peak of $A_{(-,-)}$ ($A_{(+,+)}$) is located above (below) the chemical potential. The plots (Fig.~\ref{fig:2}c and Fig.~\ref{fig:2}d) of the density of states associated to the two bands, $DOS_{(-,-)}(\omega)=\frac{1}{N_c}\sum_q A_{(-,-)}(\boldsymbol{q},\omega)$ and $DOS_{(+,+)}(\omega)=\frac{1}{N_c} \sum_q A_{(+,+)}(\boldsymbol{q},\omega)$, furtherly clarify this picture. Indeed $DOS_{(-,-)}$ ($DOS_{(+,+)}$) (see the two insets) exhibits a peak above (below) $\mu$, at $\lambda>\lambda_c$. It indicates a strong hybridization between the quasiparticles of the bare Haldane model across the topological quantum transition. In the density of states, the Van Hove singularities and the satellite bands, stemming from the EPC, are clearly distinguishable.

\begin{figure}[thb]
  \includegraphics[width=0.99\columnwidth]{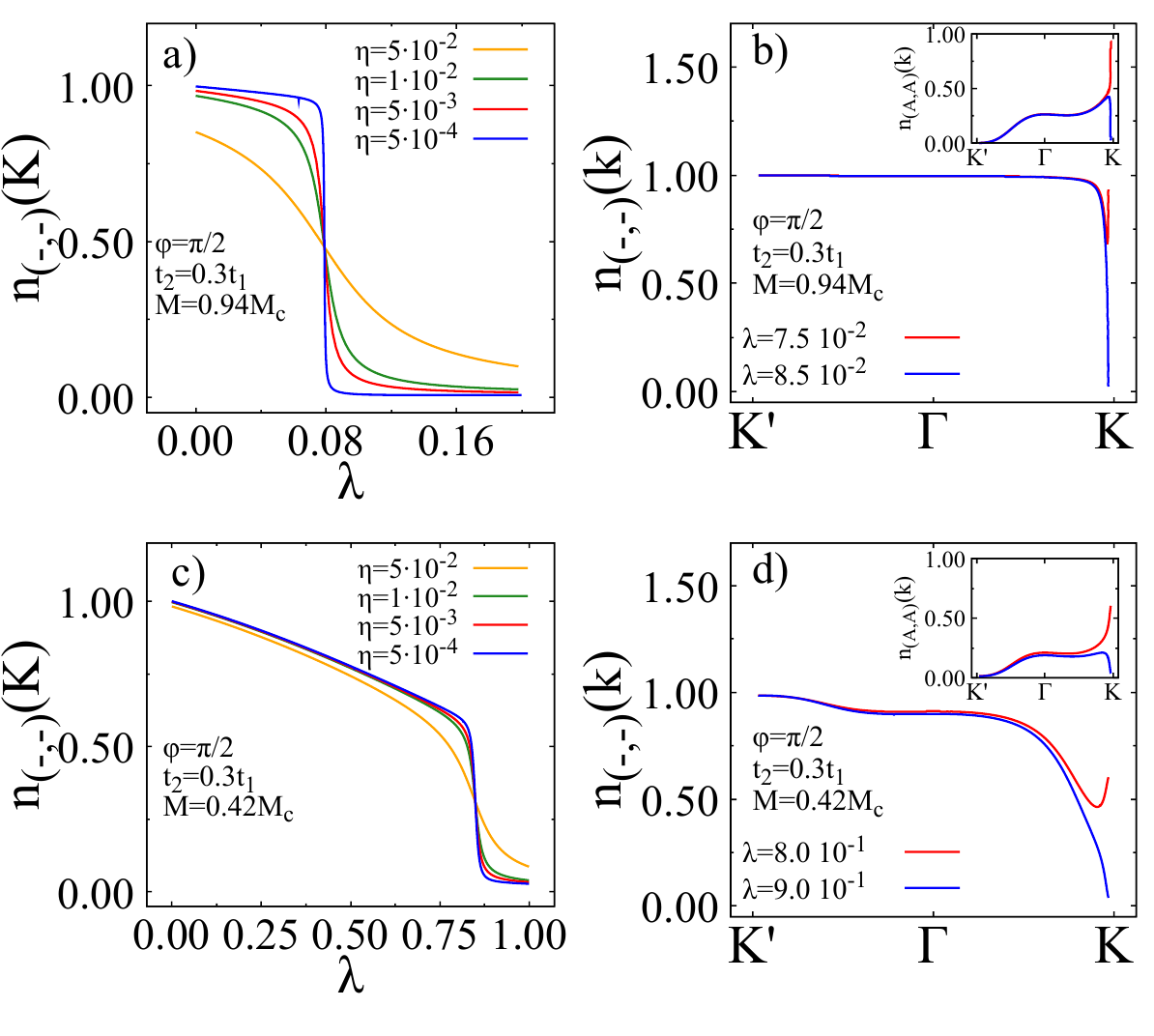}
  \caption{\label{fig:3} (color online)
    The average number of fermions $n_{(-,-)}$, at $\text{K}$ ((a) and (c)) and along $\text{K}^{\prime}-\Gamma-\text{K}$ ((b) and (d), in the inset $n_{(A,A)}$), for two different values of $M$.
    }
\end{figure}

In Fig.~\ref{fig:3}a we plot the average number of fermions $n_{(-,-)}(\boldsymbol{q})$, at $\boldsymbol{q}=\text{K}$, as function of $\lambda$: $n_{(-,-)}(\boldsymbol{q})=\int_{-\infty}^{\infty} A_{(-,-)}(\boldsymbol{q},\omega) n_F(\omega) d\omega$, where $n_F(\omega)$ is the Fermi function. By decreasing the broadening factor $\eta$, it becomes more and more clear that $n_{(-,-)}(\text{K})$ exhibits a finite discontinuity at the transition point, so that it can be used as direct indicator of the topological quantum transition. We find also (Fig.~\ref{fig:3}c) that a greater EPC is needed to destroy the topological phase when the initial parameters of the bare Haldane model are such that spinless fermions are well inside the topological phase. In this case the discontinuity at the transition point reduces indicating the presence of strong electron-electron correlations induced by the EPC. The plots in Fig.~\ref{fig:3}b and Fig.~\ref{fig:3}d, i.e the behavior, across the phase transition, of $n_{(-,-)}(\boldsymbol{q})$ along the line $\text{K}^{\prime}-\Gamma-\text{K}$, point out that the quantum transition affects only a small region of the Brillouin zone around the Dirac point $\text{K}$. 

\begin{figure}[thb]
  \includegraphics[width=0.99\columnwidth]{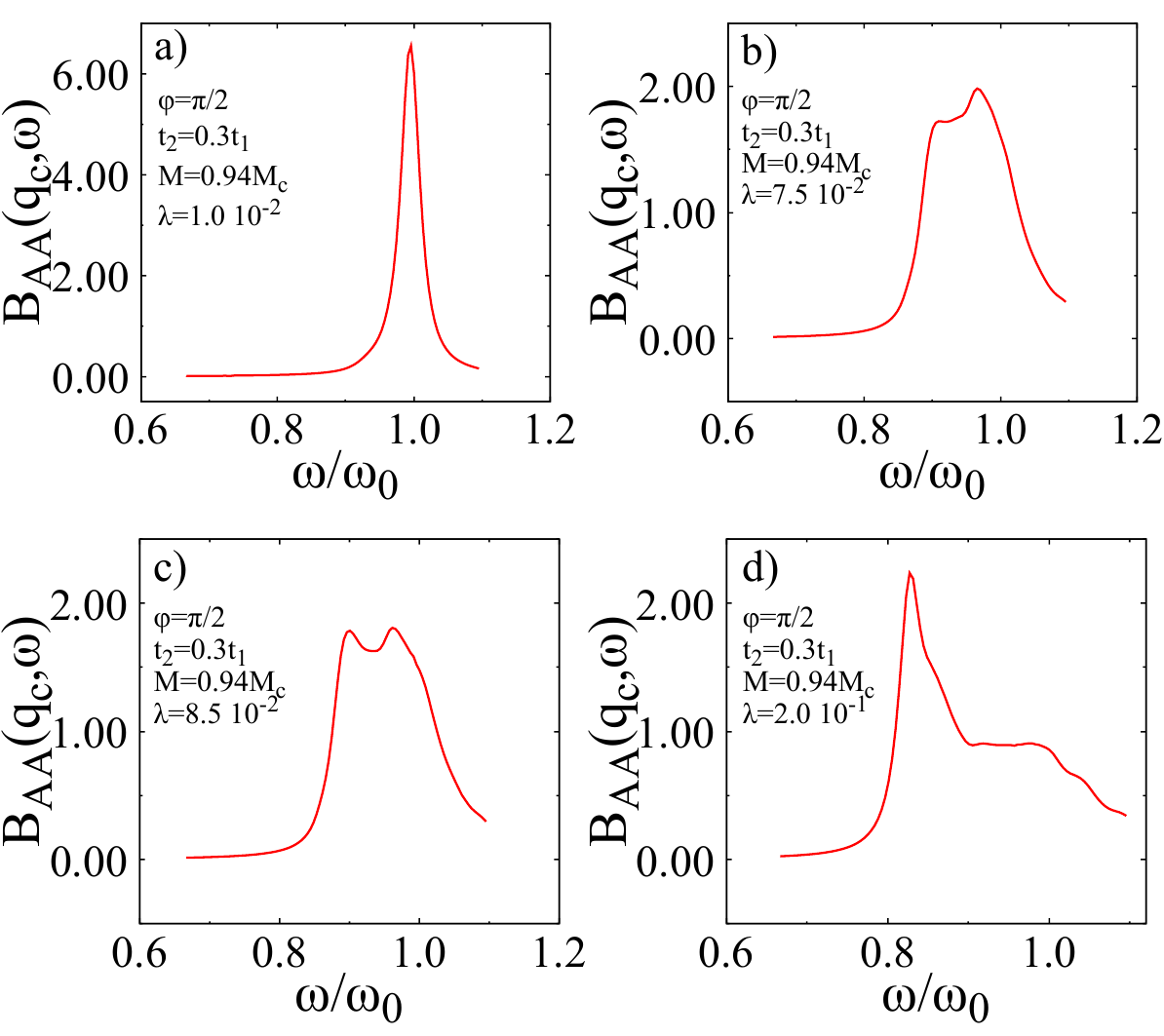}
  \caption{\label{fig:4} (color online) Phonon spectral weight function for four different values of $\lambda$. 
     }
\end{figure}
 
Finally, we investigate the effects of the quantum transition on the lattice. To this aim, we emphasize that the exact integration of the phonon degrees of freedom, through path integral technique, leads to a retarded electron-electron interaction on the same sublattice. This coupling is controlled by the bare phonon propagator, $D^0(\boldsymbol{q},z)=\frac{1}{z-\omega_0}-\frac{1}{z+\omega_0}$, and the charge-phonon vertex: $V^{0}_{i,i}(\boldsymbol{q},z)=\frac{g^2 \omega_0^2}{N_c} D^0(\boldsymbol{q},z)$ \cite{mahan, fetter}. At the lowest order in the EPC, there is not any coupling between two electrons on different sublattices, i.e. $V^{0}_{(A,B)}(\boldsymbol{q},z)=0$. On the other hand, the effective interaction between two charge carriers obeys the Dyson equation \cite{mahan, fetter}:
\begin{eqnarray}
  V^{eff}_{i,j}(\boldsymbol{q},z)=V^{0}_{i,j}(\boldsymbol{q},z)+V^{0}_{i,h}(\boldsymbol{q},z)\Pi^{*}_{h,k}(\boldsymbol{q},z\
  )V^{eff}_{k,j}(\boldsymbol{q},z),
  \nonumber
\end{eqnarray}
which defines the proper polarization insertion $\Pi^{*}_{i,j}(\boldsymbol{q},z)$. Since $\Pi^{*}_{i,j}$, in general, is a non diagonal matrix, there is an effective phonon mediated interaction even between two charge carriers located in different sublattices. At the lowest order $\Pi^{*}_{i,j}(\boldsymbol{q},z)$ is the particle-hole bubble. The next step is to replace, in this lowest order diagram, the unperturbed electron Green functions with the interacting Green functions calculated within the CPT. This procedeure allows to obtain the effective interaction between two electrons and, then, the renormalized phonon propagator $D_{i,j}$.

We focus our attention on the spectral weight function $B_{(A,A)}(\boldsymbol{q},\omega)=-\frac{\Im{D_{(A,A)}}(\boldsymbol{q},z)}{\pi}$, an odd function, that, in the absence of EPC, is peaked at $\omega=\omega_0$. At $\lambda \ne 0$, it exhibits a softening at $\boldsymbol{q_c}$ around $\frac{\boldsymbol{b_1}}{2}$ and $\frac{\boldsymbol{b_2}}{2}$, where $\boldsymbol{b_1}$ and $\boldsymbol{b_2}$ are the primitive vectors of the reciprocal lattice. In Fig.~\ref{fig:4} we plot $B_{(A,A)}(\boldsymbol{q_c},\omega)$ for four different values of the EPC. Near the topological phase transition there is a splitting of the main peak. By increasing $\lambda$, the spetral weight of the lowest (highest) energy peak increases (decreases), and, at around $\lambda_c$, the two peaks have the same intensity. We emphasize that the energy of the lowest peak is strictly related to the energy difference between the two Dirac points, both in the hole and particle sectors (compare energy of K and K' in Fig. \ref{fig:1}d and its inset), i.e. $\text{K}$ and $\text{K}^{\prime}$ are connected by the EPC. On the other hand, the highest energy peak is reminiscent of the bare phonon frequency. Finally, at $M=0$, by increasing EPC, the peak softening becomes more and more pronounced signaling a charge density wave instability (see supplemental material \cite{Supplement}).

{\it Conclusion.}
We have investigated the effects of the interaction between the vibrational modes of the lattice and the spinless charge carriers in the Haldane model on a honeycomb lattice. We found evidence of a topological quantum transition. Starting from the topological phase in the bare Haldane model, the increasing of the strength of the EPC, $\lambda$, drives the system towards a trivial insulator. By varying $\lambda$, the energy gap first decreases, closes at the transition point, and then increases. Across the transition point, a strong hybridization between the quasiparticles of the bare Haldane model occurs near the Dirac point characterized by the lowest gap. The average number of fermions exhibits a finite discontinuity at the transition in this particular point of the Brillouin zone and can be used as direct indicator of the topological quantum transition. We have also shown that the renormalized phonon propagator exhibits a two peak structure across the quantum transition, whereas, in absence of the mass term, there is indication of a complete softening of the effective vibrational mode signaling a charge density wave instability.




%
%

\end{document}